\documentclass[useAMS,usenatbib]{mn2e}
\usepackage{graphicx}
\usepackage{setspace}
\usepackage{natbib}
\usepackage{color}
\usepackage{amsmath,amssymb}
\usepackage{times}
\usepackage{aas_macros}

\title{Exposing Sgr tidal debris behind the Galactic disk with
  M giants selected in WISE$\cap$2MASS}

\author[Koposov et al.]{S. E. Koposov$^{1}$
 \thanks{E-mail:koposov@ast.cam.ac.uk,vasily@ast.cam.ac.uk}, 
V. Belokurov$^{1}$, 
D. B. Zucker$^{2,3,4}$,
G. F. Lewis$^{5}$,
R. A. Ibata$^{6}$,\newauthor
E. W. Olszewski$^{7}$,
\'A. R. L\'opez-S\'anchez$^{4,3}$,
E. A. Hyde$^{2,3}$
\\ $^{1}$Institute of Astronomy, Madingley Rd, Cambridge, CB3 0HA, UK
\\ $^{2}$Macquarie University Research Centre in Astronomy, Astrophysics \& Astrophotonics, NSW 2109, Australia
\\ $^{3}$Department of Physics and Astronomy, Macquarie University, North Ryde, NSW 2109, Australia
\\ $^{4}$Australian Astronomical Observatory, PO Box 915, North Ryde, NSW 1670, Australia
\\ $^{5}$Sydney Institute for Astronomy, School of Physics, A28, University of Sydney, Sydney NSW 2006, Australia
\\ $^{6}$Observatoire Astronomique de Strasbourg, Universit\'e de Strasbourg, CNRS, UMR 7550, 11 rue de l'Universit\'e, F-67000 Strasbourg, France
\\ $^{7}$Steward Observatory, University of Arizona, Tucson, AZ 85721, USA
}

\date{\today}
\def\change#1{{#1}}
\begin{document}
\maketitle

\begin{abstract}

We show that a combination of infrared photometry from WISE and 2MASS
surveys can yield highly pure samples of M giant stars. We take
advantage of the new WISE$\cap$2MASS M giant selection to trace the
Sagittarius trailing tail behind the Galactic disk in the direction of
the anti-centre. The M giant candidates selected via broad-band
photometry are confirmed spectroscopically using AAOmega on the AAT in
3 fields around the extremity of the Sgr trailing tail in the Southern
Galactic hemisphere. We demonstrate that at the Sgr longitude $\tilde
\Lambda_{\odot} = 204^{\circ}$, the line-of-sight velocity of the
trailing tail starts to deviate from the track of the \citet{law10}
model, confirming the prediction of \citet{belokurov14}. This
discovery serves to substantiate the measurement of low differential
orbital precession of the Sgr stream which in turn may imply diminished
dark matter content within 100 kpc.
\end{abstract}

\section{Introduction}

Across the Milky Way, the safest place to conceal the remains of a
destroyed satellite is behind the Galactic disk. This is exactly where
the third largest, and the closest, companion galaxy, the Sagittarius
dwarf, avoided discovery until the end of the last millennium
\citep{Sgrdiscovery}. In the twenty years that followed, through the
meticulous identification of particular stellar tracers, Sagittarius's
enormous tidal tails have been shown to wrap around the entire Milky
Way \citep{Mateo1996, Totten1998, Mateo1998, Majewski1999, Ivezic2000,
Yanny2000, ibata01, MD2001, Vivas2001, Dohm2001, Newberg2002, Newberg2003,
majewski03, Belokurov2006, Yanny2009, NO2010,
Correnti2010,Ruhland2011,koposov12,belokurov14}. Yet, after two
decades of collaborative effort, two minute gaps are still present:
close to the Galactic plane, no choice of stellar tracer results in a
selection clean enough not to be completely swamped by the nearby disk
dwarfs. As a result, in the direction towards the bulge, the base of
the leading tail is lacking, while in the opposite direction, towards
the Galactic anti-centre, the sight of the trailing debris is lost as
it reaches into the disk from below the plane.

\begin{figure*}
\includegraphics[width=0.95\textwidth]{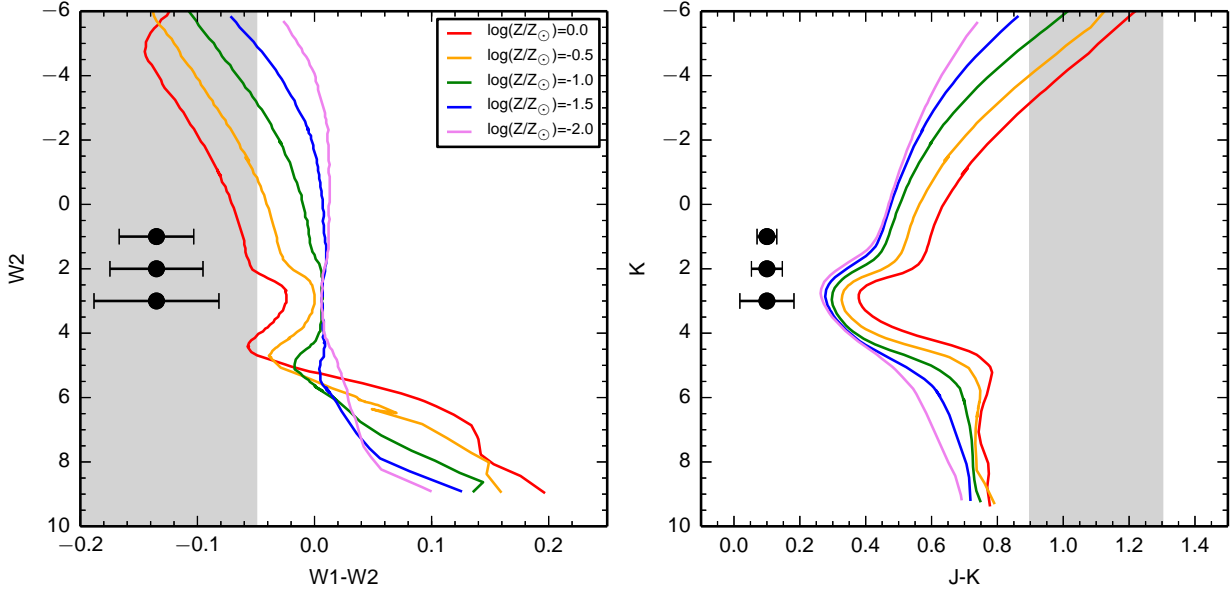}
\caption{PARSEC isochrones for an old stellar population with five
  different metallicities in infrared pass-bands. {\it Left:}
  Isochrones in WISE W1, W2 bands. {\it Right:} Isochrones in 2MASS
  J,K filters. Note the change in the behaviour of the metal-rich RGB
  stars: while being red in the near-IR (and optical), these are blue
  in the WISE colours. Grey shows the colour range used for the M
  giant selection proposed here. \change{The horizontal error-bars
    show the approximate photometric errors expected for the colours of
    an M giant with absolute magnitude in $M_{W2}\sim$ 5 at distances
    of 20, 40 and 60\,kpc.}}
\label{fig:iso}
\end{figure*}

These short (at most 20$-$30 degrees on the sky) missing stream
pieces, nevertheless, hold tantalising clues to the disruption of the
Sagittarius dwarf (Sgr). If it were not obscured by the disk and the
bulge, the view of the area where the leading tail attaches to the
remnant's body could help explain the peculiar stream bifurcation. As
pointed out by \citet{Belokurov2006}, the leading tail appears
bifurcated into the bright and faint components. As of today, no
convincing mechanism has been found to produce such splitting of the
Sgr tidal tails. Several hypotheses have been brought up: i) a group
or binary in-fall \citep[e.g.][]{Helmi2011,koposov12}, and ii) a
rotating progenitor \citep{Jorge2010,Jorge2011}. Each of these
scenarios predicts distinctly different behaviour of the leading
tail(s) in the vicinity of the progenitor.

Similarly, as pointed out by \citet{belokurov14}, the exact distance
and velocity of the trailing debris around the Galactic anti-centre
can be used to understand the Sgr stream's precession. They identify new
far-flung tidal debris coincident with Sgr's orbital plane in the
Northern hemisphere, and argue that, having crossed the Milky Way's
disk, the trailing tail reaches much larger distances than
expected. In so doing, it attains its maximal extent, or apo-centre,
later than predicted by most current models, yielding, therefore, a
smaller orbital precession angle. The larger apo-centric distance and
the smaller differential precession point towards a significantly
lower dark matter content within the stream's orbit, as has been
recently demonstrated by \citet{Gibbons2014}. However,
\citet{belokurov14} based their inference on the assumption that the
distant stream detected above the disk in the North is simply the
continuation of Sgr's Southern trailing tail. While this
conjecture might seem reasonable, as both streams separated by the
Galactic disk follow very similar distance and velocity gradients, it
is at odds with established models of Sgr's disruption
\citep[see, e.g.,][]{law10}.

\change{The most tangible difference between the two pictures of 
  Sgr trailing arm behaviour around the Galactic anti-centre is
  apparent in Figure~11 of \citet{belokurov14}. This shows the
  evolution of the line-of-sight velocities of the Sgr tails as a
  function of the Sgr longitude, $\tilde \Lambda_{\odot}$. While at
  most longitudes the data and the model are in good agreement, the larger
  apocentric distance of 100\, kpc observed by \citet{belokurov14}
  implies that, on approach to the Galactic disk, the Sgr trailing
  tail should deviate significantly from the model of
  \citet{law10}. Namely, the radial velocity of the trailing debris is
  predicted to be appreciably faster (higher negative velocity) than
  that of the model. Thus the key test of the \citet{belokurov14}
  conjecture is the measurement of the stream's line-of-sight
  velocities near the Galactic plane, where the stream stars return
  from the apocenter. In Sgr stream coordinates, the region of
  interest is around $\tilde \Lambda_{\odot} \sim 200$, which
  corresponds to a rather low Galactic latitude $|b| \lesssim 20^{\circ}$.}


%
\begin{figure*}
\includegraphics[width=0.95\textwidth]{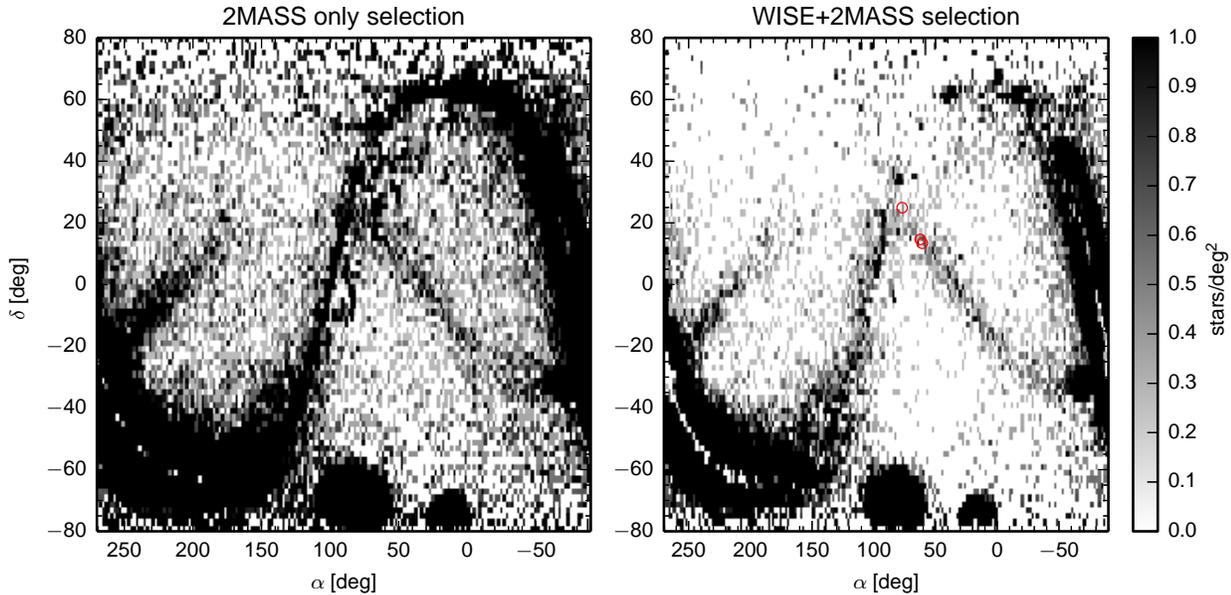}
\caption{{\it Left:} Density map of the M giant stars selected using
  2MASS photometry only, i.e. $0.9 < {\rm J} - {\rm K} < 1.3$,
  $0.22<{ \rm J} - {\rm H} -0.561\,( {\rm J} - { \rm K })<0.46$ and
  $11<{\rm K}<13$ \citep[see also][]{majewski03}. {\it Right:} Density
  map of the M giants selected using WISE$\cap$2MASS: $0.9<{\rm J} -
  {\rm K} < 1.3$, $-0.2<{\rm W1}-{\rm W2}<-0.05$ and $11<{\rm
    K}<13$. Note the cleaner M giant selection provided by the
  combination of WISE and 2MASS, at both low and high Galactic
  latitudes. \change{The locations of three fields targeted with
    AAOmega spectroscopy are indicated with open red circles.}}
\label{mgiantsmap}
\end{figure*}
\begin{figure*}
\includegraphics[width=0.95\textwidth]{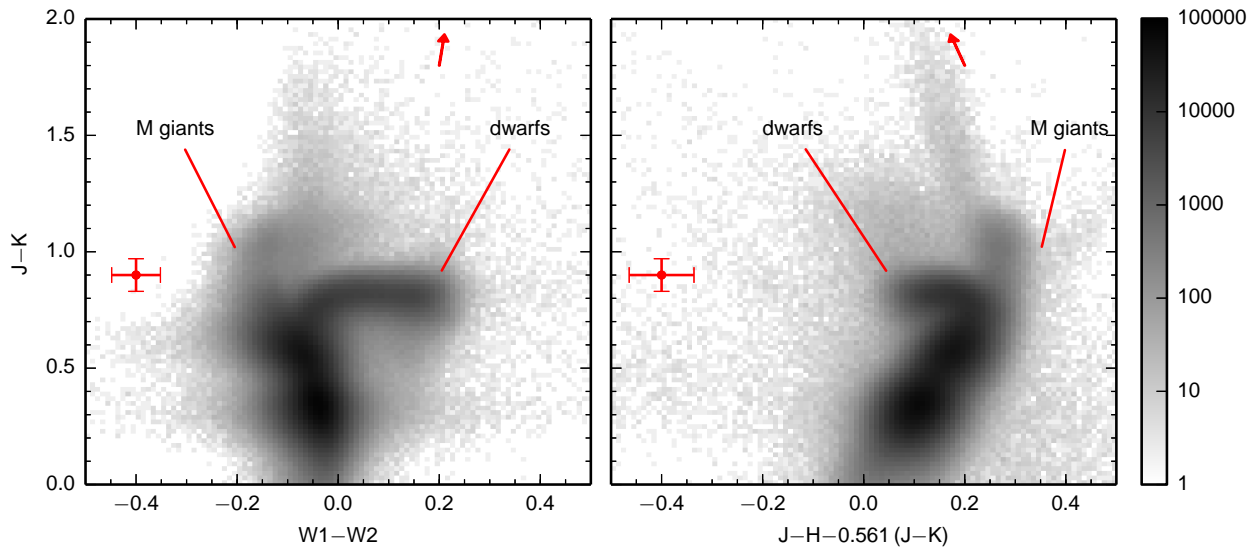}
\caption{\change{Stellar density distribution in near-IR and IR
    colour-colour spaces.  {\it Left:} Positions of stars
    with $J\sim 13$ and $|b|>10$ in the (W1$-$W2, J$-$K) plane. The
    dwarf and giant sequences start to separate at W1$-$W2$\sim -0.1$
    and J$-$K$\sim0.8$. {\it Right:} The same stars are now plotted in
    the (J$-$H$-$0.561\,(J$-$K), J$-$K) space. The X-axis of this
    panel is the colour combination used for M giant selection by
    \citet{majewski03}. In both panels, small arrows show the
    direction of the dust extinction, while the error-bars illustrate
    the characteristic photometric uncertainty for an M-giant with a
    magnitude of J$\sim$ 14. In the WISE bands, the bulk of metal-rich
    M giants (at negative W1$-$W2) is further away from the dwarf
    population as compared to the 2MASS JHK color combination (at
    similar J$-$K color). We argue that, at faint magnitudes, this
    leads to noticeably cleaner samples of M giants.} }
\label{fig:colcol}
\end{figure*}

Previously, a wide variety of stellar types have been used to trace
 Sgr tidal debris.  For example, main sequence (MS) and main
sequence turn-off (MSTO) stars are the most numerous amongst the Sgr
populations, and have been used with great success to chart the
position of the tails on the sky \citep[e.g.][]{koposov12}. However,
at distances beyond 30\,kpc, these are too faint for spectroscopic
follow-up.  While blue horizontal branch (BHB) stars are perfect
distance indicators, their density in this part of the trailing tail
is surprisingly low \citep[see][]{koposov12}; more importantly, in the
vicinity of the Galactic disk, the efficiency of BHB identification
quickly deteriorates with mounting dust reddening.  Red giants are
bright spectroscopic targets, but even at high latitudes
photometrically selected candidates are dominated by the thick disk
dwarfs. Therefore, their follow-up close to the Milky Way's plane
appears impractical.  The two remaining tracers that might withstand
significant levels of disk dwarf contamination include RR Lyrae
and M giants. RR Lyrae stand apart from any other stellar population
due to their characteristic variability. For instance, the OGLE survey
has produced multi-epoch datasets in which RR Lyrae are routinely
detected beyond the bulge and the disk at latitudes as low as
$|b|\sim2^{\circ}$ \citep[see e.g.][]{OGLE2012}. Unfortunately, the
region of interest around the Galactic anti-centre has not been
surveyed with sufficient temporal resolution and depth to identify RR
Lyrae in the Sgr stream.

Taking advantage of the full-sky coverage of the 2MASS survey
\citep{2mass}, \citet{Ibata2002} and \citet{majewski03} applied the
infrared selection criteria outlined by \citet{Bessell1988} to pick
out distant halo M giants. As these studies demonstrate, at high
latitudes and distances beyond 20\,kpc, Sgr debris dominates the M giant
counts. This ought to be caused by a combination of three distinct
effects: the typical lifetime of an M giant star, the star-formation
history of Sgr and the paucity of recent massive dwarf galaxy accretion
onto the Milky Way. It turns out that, within 30$-$40\,kpc from the
Sun, the M giant selection based on 2MASS $JHK$ magnitudes can
distinguish giants from dwarfs quite efficiently, even at Galactic latitudes
as low as $|b|\sim20^{\circ}$, as illustrated by the discovery of the
debris from an accretion event passing through the constellations of
Triangulum and Andromeda \citep{Rocha2004}. However, for fainter stars
and/or objects closer to the Galactic plane, M giant selection
based on 2MASS photometry starts to falter. Therefore, to be able
to proceed with the kinematic study of the Sgr trailing tail around
the anti-centre, we strive to optimise M giant identification by
complementing the 2MASS photometry with far-infrared data from the
WISE survey \citep{wise}. Section 2 gives the details of the proposed
tracer selection, while Section 3 presents the results of the
spectroscopic follow-up of the Sgr M giant candidate stars.

\section{M giant selection with WISE$\cap$2MASS}

Figure~3 of \citet{majewski03} exemplifies the power of
infrared-based M giant selection: this was the very first time the
immense scale of Sgr's disruption was shown in such
clarity. However, in this Figure the contamination of the M giant
sample at low latitudes is already obvious, even though the most
affected areas with $|b|\lesssim 15^{\circ}$ have been excised. It is
clear that the original identification scheme breaks down at i) large
distances and ii) high reddening as the limits of the 2MASS photometry
are approached. Additionally, next to the Milky Way plane, the number
of false positives due to disk dwarfs grows overwhelmingly quickly making
a spectroscopic follow-up campaign too costly. Below we show that these
problems can be remedied if, for selecting Sgr stream M giants,
2MASS data are used in conjunction with far-infrared photometry from the
WISE survey.

\begin{figure}
\includegraphics{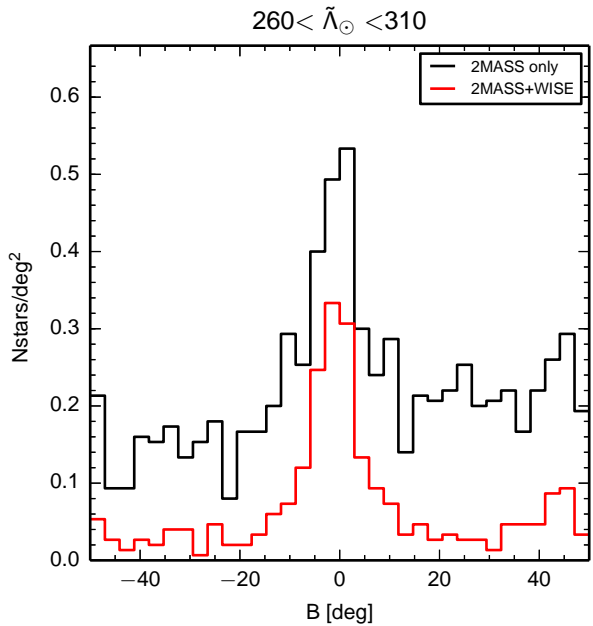}
\caption{Density profiles of the Sgr trailing tail in the range 
  $260^{\circ} < \tilde \Lambda_{\odot} < 310^{\circ}$ as traced by M
  giants selected using 2MASS only (black) and WISE$\cap$2MASS
  (red). While the stream signal remains largely the same in both
  profiles, the foreground contamination appears dramatically reduced
  when WISE$\cap$2MASS is employed.
 \label{fig:profile}}
\end{figure}

Intriguingly, the WISE data turn out to be extremely valuable for the
particular task of identifying giant stars. The information
gain here is due to the peculiar behaviour of the mid-infrared colours
of metal-rich giants. Optical and near-infrared broad-band colours of
giant stars are quite similar to those of dwarfs, making 
giant tracer selection rather inefficient. Conversely, in WISE
photometry, metal-rich giants stand out from dwarfs thanks to the
presence of the gravity-sensitive CO-bands. Unusually, this makes the
metal-rich giants {\it bluer} in W1$-$W2 colours compared to dwarf stars!

\begin{figure*}
\includegraphics[width=0.95\textwidth]{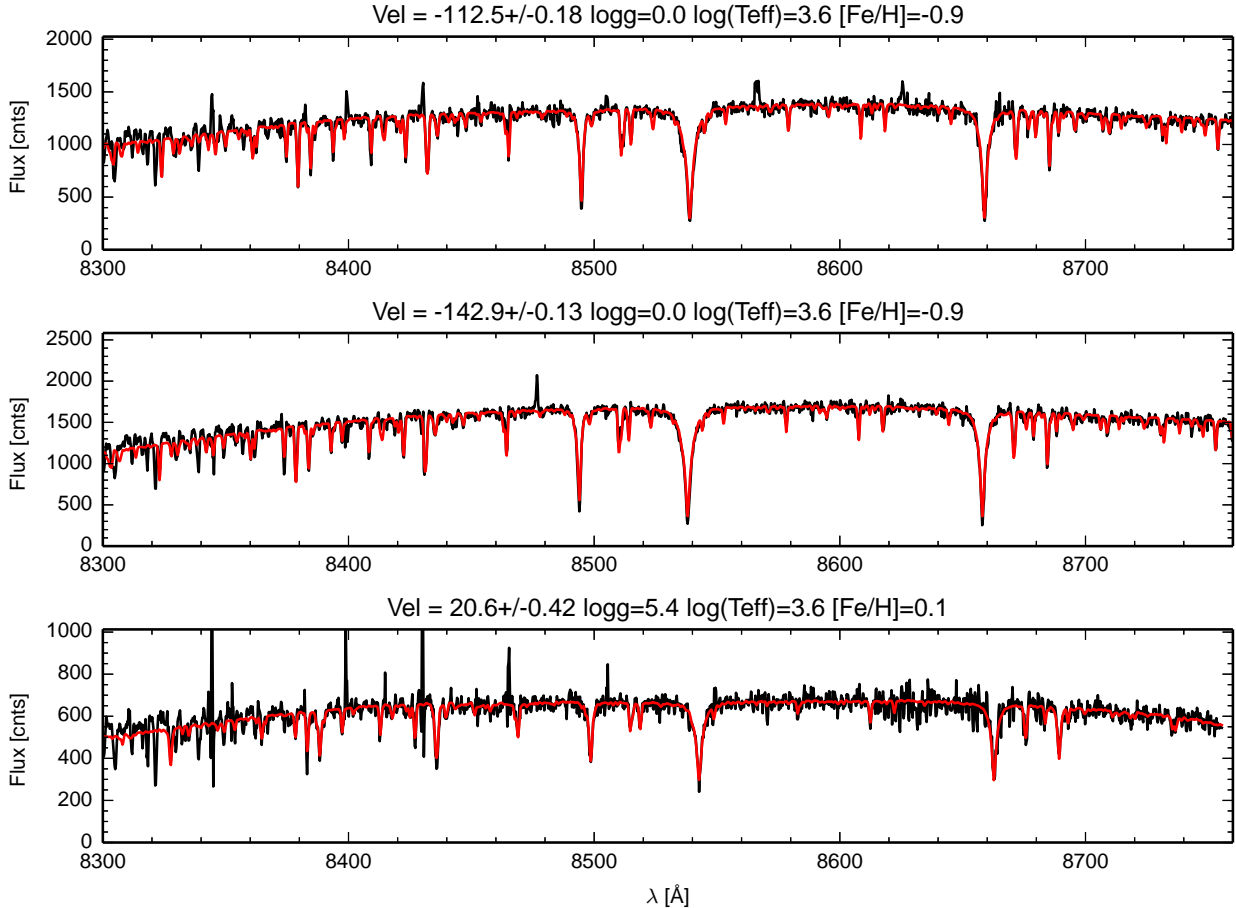}
\caption{Sample AAOmega spectra of three of our M giant
  candidates. Black shows the data, red lines show the best-fit model
  templates. The spectra in the top and middle panels correspond to
  Sgr stars while the bottom spectrum is a foreground dwarf star.}
\label{fig:spectra}
\end{figure*}

Figure~\ref{fig:iso} illustrates this effect. Here, five 12 Gyr PARSEC
\citep{parsec} isochrones with different metallicities are shown. The
near-IR isochrones look as usual, i.e., the metal-rich tracks are
redder than metal-poor tracks. Moreover, both giants and dwarfs are
also redder compared to the turnoff stars. However, along mid-IR
isochrones, the metal-rich giant branch actually turns blue in the
W1$-$W2 colour, which would clearly make the M giants stand out
\change{further from the dwarfs}. This behaviour of the mid-IR giant
tracks points out the route for their efficient selection: we need
stars that are red in the 2MASS bands, $({\rm J}-{\rm K})\gtrsim 0.9$, and blue in
the WISE bands, ${\rm W1}-{\rm W2}\lesssim -0.05$. \change{Additionally,
  Figure~\ref{fig:iso} shows the expected median photometric
  uncertainty for an M-giant with ${\rm M_{W2}}=5$ placed at heliocentric
  distances of 20, 40 and 60\,kpc. This demonstrates how the photometric
  accuracy deteriorates as stars at larger distances are selected.}
As far as distant halo tracers or areas with large levels of
extinction are concerned, the efficiency of this giant identification
is limited by the quality of the WISE photometry. At around
W1$\sim$14, the WISE photometric errors become large enough that the
candidate tracers are swamped by the dwarf stars with
${\rm W1}-{\rm W2}\gtrsim 0$.

Figure~\ref{mgiantsmap} demonstrates the improvement in the quality of
the M giant selection. It compares two all-sky density maps: one (left
panel) obtained by applying the original 2MASS selection as proposed
by \citet{majewski03}, and another (right panel) built using the
following combination of WISE and 2MASS:

\begin{eqnarray}
-0.2 < ({\rm W1}-{\rm W2})_0 < -0.05\nonumber\\
11 < {\rm W1_0} < 13.5\\
0.9< ({\rm J}-{\rm K})_{2MASS,0} < 1.3 \nonumber
\end{eqnarray}

\noindent Here, all magnitudes are corrected for extinction using the
maps provided by \citet{dust}.

\begin{figure}
\includegraphics{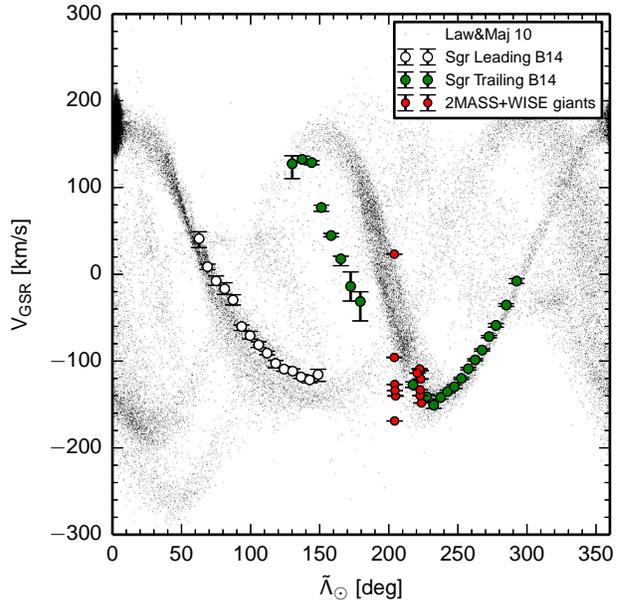}
\caption{Galactocentric radial velocities of the Sgr tidal debris as a
  function of the angle along the stream. Small black dots show the
  phase-space positions of the N-body particles in the \citet{law10}
  model. Green and unfilled circles with error-bars are the
  compilation of the available kinematic data for the trailing and
  leading tail, respectively, from \citet{belokurov14}. Filled red
  circles show the new M giant radial velocities reported in this
  work. Note that our data agree with the literature values at $\tilde
  \Lambda_{\odot} \sim 220^{\circ}$. However, closer to the disk, at
  $\tilde \Lambda_{\odot} = 204^{\circ}$, we show that the trailing
  tail is at lower velocity than predicted by the \citet{law10}
  model. This finding supports the hypothesis of \citet{belokurov14}
  as to the properties of the trailing tail in the Northern
  hemisphere. \label{fig:kinematics}}
\end{figure}

The two density distributions have features in common. For example,
the leading ($100^{\circ}<\alpha<260^{\circ}$) as well
as the trailing tail ($-60^{\circ}<\alpha<80^{\circ}$) of the Sgr stream, are equally
prominent in both maps, although perhaps in the right panel there is
a minor depletion in M giant counts in the most distant part of the
leading tail, where WISE photometry starts to deteriorate. However,
there are more important differences too. The halo, i.e., the portions
of the sky far away from the Galactic disk, appears both tidier and
emptier with the new selection. Also, there is a lot less junk close
to the disk, while the disk itself appears significantly reduced in
density, especially around the anti-centre - the region of interest
($\alpha\sim 100^{\circ}$).  In addition, it is worth noting that the
new WISE$\cap$2MASS M giant map also boasts a clearer view of the
Triangulum-Andrometa merger (or Tri-And, see Deason et al 2014 for
details ) at $-20^{\circ}\lesssim \alpha \lesssim 80^{\circ}$ and
$10^{\circ} \lesssim \delta \lesssim 50^{\circ}$. 

Figure~\ref{fig:profile} gives a more quantitative summary of the
efficiency of the new selection. The Figure presents two density
profiles of the same piece of the Sgr trailing tail with $260^{\circ}
< \tilde \Lambda_{\odot} < 310^{\circ}$: one produced using 2MASS only
(black), and one employing the WISE$\cap$2MASS selections (red). As
demonstrated here, the amplitude of the stream detection remains
largely unchanged when switching to WISE$\cap$2MASS, while the
foreground contamination is significantly reduced. \change{By fitting
  the Gaussian model together with the linearly changing foreground
  density to both datasets we find that the efficiency of selecting
  Sgr M-giant stars using WISE$\cap$2MASS is within 10\% of
  the efficiency of the 2MASS-only selection, while the contamination
  is $\sim$ 6 times lower for this particular area of the sky.
  Overall, given the performance of the new M giant selection for the
  detection of previously identified stellar halo sub-structures,
  we conclude that the WISE photometry indeed helps to reduce the
  contamination levels appreciably.}

\change{Given that the quality of any "colour-cut" selection method is
  a complicated function of i) the distribution of the sources of
  interest in colour-space, ii) the distribution of their
  contaminants, as well as iii) the photometric errors and iv)
  extinction coefficients, it is not straightforward to pinpoint one
  factor responsible for the increased efficiency of the proposed
  WISE$\cap$2MASS selection. Nonetheless, Figure~\ref{fig:colcol}
  provides a further exposition as to why the WISE photometry might be
  of help. This Figure compares the behaviour of stellar loci in the
  (W1$-$W2, J$-$K) space to that in (J$-$H$-$0.561\,(J$-$K), J$-$K),
  the space originally used by \citet{majewski03}. We argue, that due
  to the peculiar behaviour of the WISE colours of the metal-rich
  giants, these stars tend be further apart from the bulk of the
  dwarfs (their main contaminants) at similar J$-$K
  colour. Furthermore, as illustrated by the error-bars over-plotted,
  at fainter magnitudes J$\sim$14, the 2MASS-only selection is more
  affected by photometric errors. This photometric deterioration is
  exacerbated in the presence of significant extinction; the
  2MASS-only selections will suffer a more pronounced loss due to
  larger extinction coefficients.}

\begin{table}
\caption{Location of the 2dF fields}
\begin{tabular}{ccccccc}
Field & $\tilde\Lambda_{\odot}$ & $\tilde B_{\odot}$ & $\alpha$ & $\delta$ & $l$ & $b$ \\
 & deg & deg & deg & deg & deg & deg \\
\hline
1 & 204.2 &  2.7 & 77.1 & 24.8 & 178.8 & -9.2  \\
2 & 221.0 & -1.0 & 62.4 & 14.6 & 178.2 & -26.4 \\
3 & 223.0 & -1.4 & 60.6 & 13.3 & 178.0 & -28.5 \\
\end{tabular}
\label{tab:areas}
\end{table}

\section{Kinematics of the Sgr trailing tail around the anti-centre}

\begin{table*}
\caption{Kinematic and photometric properties of M giant candidates}
\begin{tabular}{rlrrrrrrrrr}
\hline
  ID & Field   &      $\alpha$ &     $\delta$ &   V$_{GSR}$ &   $\sigma_V$ &    W1 &    W2 &   J(2MASS) &   K(2MASS) &   log(g) \\
   &    &    deg &     deg &   km\,$s^{-1}$ &   km\,$s^{-1}$ &    mag &    mag &   mag &   mag &  dex \\

\hline
     0 & 1  & 77.3832 & 24.7543 &        15.3 &          0.4 & 13.43 & 13.47 &      14.65 &      13.49 &    5.4 \\
     1 & 1  & 77.0267 & 24.75   &      -126.9 &          0.2 & 13.15 & 13.23 &      14.64 &      13.34 &    0   \\
     2 & 1  & 76.5295 & 24.2549 &      -140.1 &          0.3 & 12.75 & 12.83 &      14.08 &      12.82 &    0   \\
     3 & 1  & 76.5367 & 25.3871 &      -133.9 &          0.3 & 12.35 & 12.41 &      14.1  &      12.57 &    0   \\
     4 & 1  & 77.0282 & 24.9731 &      -169   &          0   & 11.63 & 11.75 &      13.26 &      11.95 &    0   \\
     5 & 1  & 77.0944 & 25.6639 &      -112.5 &          1   & 13.65 & 13.79 &      15.2  &      13.82 &    5.6 \\
     6 & 1  & 77.1975 & 25.1967 &        23.3 &          0.2 & 11.57 & 11.68 &      12.93 &      11.7  &    0   \\
     7 & 1  & 77.1774 & 25.0204 &       -95.8 &          0.3 & 13.21 & 13.27 &      14.47 &      13.27 &    0   \\
     8 & 2  & 61.1325 & 13.2303 &         0.4 &          0.3 & 13.08 & 13.14 &      14.3  &      13.2  &    5.5 \\
     9 & 2  & 60.4733 & 12.4608 &      -148.3 &          0.1 & 11.78 & 11.83 &      13.1  &      11.91 &    0   \\
    10 & 2  & 60.2445 & 13.874  &      -120.7 &          0.2 & 12.61 & 12.69 &      13.86 &      12.66 &    0   \\
    11 & 2  & 60.5928 & 13.5688 &      -110.9 &          0.3 & 13.5 & 13.56 &      14.88 &      13.77 &    0   \\
    12 & 2  & 61.1166 & 14.1835 &      -109.1 &          0.3 & 13.35 & 13.44 &      14.66 &      13.43 &    0   \\
    13 & 2  & 61.0215 & 13.6043 &      -139.8 &          0.1 & 11.89 & 11.94 &      13.16 &      12.09 &    0   \\
    14 & 2  & 61.0472 & 13.4434 &      -133.2 &          0.2 & 12.22 & 12.28 &      13.52 &      12.45 &    0   \\
    15 & 3  & 62.5388 & 15.3289 &      -113.9 &          0.2 & 12.38 & 12.44 &      13.83 &      12.58 &    0   \\
\hline
\end{tabular}
\label{tab:rvs}
\end{table*}

We used the newly devised method of M giant identification, as
described in the previous section, to pick out targets for a
spectroscopic follow-up program. To measure the line-of-sight
velocities of Sgr debris, 2dF+AAOmega on Anglo-Australian Telescope (AAT) was used to target 3
locations along the trailing tail.\footnote{We originally planned to
  observe 5 fields extending all the way to the disk plane, however
  only 3 were eventually observed.} The exact coordinates were chosen
according to a small number of simple rules. First, two validation
fields (Fields 2 and 3) were selected, i.e., locations where Sgr
debris has been previously studied with spectroscopy. Second, a field
(Field 1) as far away along the trailing tail from the last known
debris detection was identified.
When placing the fields we attempted to locate them in such a way as to maximise
the number of M giant star candidates in the 2dF field of view as well as to avoid 
very extincted patches of the sky. The coordinates of all three
pointings are listed in Table~\ref{tab:areas}. To fill all the AAOmega
fibers, the WISE$\cap$2MASS M giant candidate sample was
complemented by additional objects having colours and magnitudes
consistent with the red giant branch stars at distances $30\lesssim
R_{hel} \lesssim 100\,$kpc. Note that one of the fields was
selected to lie within the  UKIDSS GCS \citep{lawrence07} survey footprint, hence we used
 UKIDSS J, K photometry instead of 2MASS. To summarise, Field 1
had 8 M giant candidates allocated, Field 2 had 7, and, finally, Field
3 had only 1 WISE$\cap$2MASS M giant candidate, yielding a total of 16 
across all 3 fields.

The designated three fields in the Sgr Stream near $l,b = (180, -20)$
were observed with the 2dF top end $+$ AAOmega spectrograph on the
3.9m Anglo-Australian Telescope the night of 27 November, 2013, as
part of service programme AO179. 2dF is a robotic positioner which
allocates up to 392 $2^{\prime\prime}$ fibres to targets across a
2$^{\circ}$ diameter field of view. These fibres feed AAOmega's blue
and red arms simultaneously, which for our observations were
configured with 580V and 1700D gratings for central wavelengths /
resolutions of 4800\AA \,/ 1300 and 8540\AA \,/ 8000,
respectively. Each field was observed for $3 \times 1800$s, at
airmasses ranging from $\sim 1.4$ -- $2$ and typical seeing of $\sim
2.5^{\prime\prime}$, with quartz spectroscopic fibre flat fields and
arc lamp spectra taken immediately prior to each set of exposures. We
processed both blue and red camera data using version 5.53 of the
2dfdr reduction pipeline, although in this work we consider only the
red spectra, i.e., those centred on the Ca II infrared triplet.

Following the approach described in detail in \citet{Koposov2011}, the
reduced spectra were then analyzed to determine the best-fit stellar
template, chosen from the PHOENIX grid \citep[see][]{phoenix}, whilst
simultaneously measuring line-of-sight velocities. All objects with
the best-fit $\log(g)< 2$ were classified as giants. Among the 16
stars targeted as M giant candidates, 13 had $\log(g)<2$ (6 from Field
1, 6 from Field 2 and 1 from Field 3). Example spectra of three of
the observed stars are shown in Figure~\ref{fig:spectra} together with
their best fitting models.  According to the results of our fitting
procedure, of these three, two are giants (top and middle) and one is a
dwarf (bottom). Notwithstanding the relatively small sample size, our
spectroscopic follow-up program confirms the high efficiency (in excess of
80\%) of the WISE$\cap$2MASS M giant selection method.

Figure ~\ref{fig:kinematics} gives the summary of all M giants with
measured radial velocities (corrected for the solar motion) in our
AAOmega spectroscopic sample (red circles) as a function of the angle
along the stream, $\tilde \Lambda_{\odot}$. The Figure also shows
previous measurements of the line-of-sight velocities of the leading
(empty circles) and the trailing (green circles) tails as well as
the prediction from the model of \citet{law10} (black
dots). Reassuringly, the radial velocities of the Sgr trailing tail in
the two validation fields (at around $\tilde \Lambda_{\odot} \sim
220^{\circ}$) match well the values from the literature. Stepping
farther from the progenitor along the trailing tail, Field 1 at
$\tilde \Lambda_{\odot} \sim 204^{\circ}$ shows a clear deviation from
the kinematics stipulated by the N-body simulation. Here, 5 out of 6
confirmed M giants are at lower radial velocities, following the trend
suggested by \citet{belokurov14}. The sixth M giant star has $V_{\rm
  GSR} \sim 20$ km s$^{-1}$. We surmise that this is unlikely to be a
dwarf interloper (even though this field is at a lower Galactic
latitude than the other two) but rather a member of the Tri-And
structure \cite[see the recent studies of][]{Deason2014,
  Sheffield2014}. 
  
\begin{figure}
\includegraphics{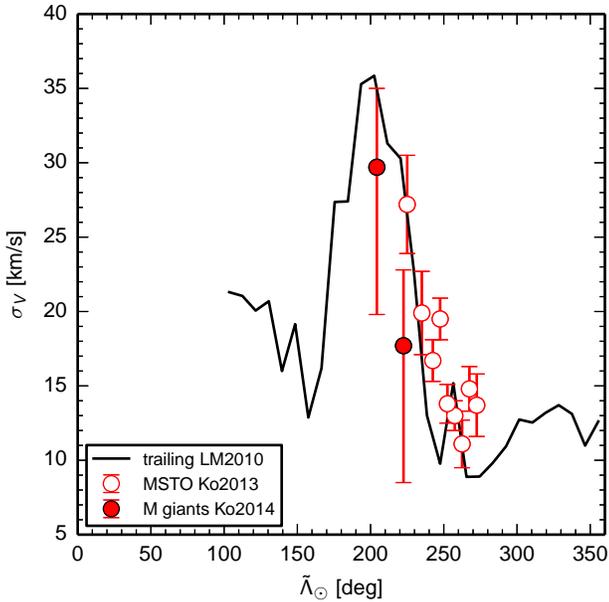}
\caption{The velocity dispersion of the Sgr Stream trailing tail as measured by
M giants (filled points), MSTO stars from \citet{koposov13} (empty circles) and compared with 
the predictions of the Law \& Majewski model (black solid line).}
\label{fig:veldisp} 
\end{figure}

Interestingly,  the radial velocities of M giants in the Sgr stream show an
appreciable scatter. This is not likely to be caused by the contamination or
the accuracy of our velocity measurements but  rather by the behaviour of the
tidal stream around the turn-around in the phase space. A similar trend of
increasing stream dispersion was in fact observed before, for example in
\citet{koposov13}. This is, of course, also predicted by Sgr disruption
models in general and the Law \& Majewski model in particular. Figure~\ref{fig:veldisp} shows how the velocity dispersion that we derive from M giants compares to
the prediction of the Law \& Majewski model as well with the measurements by
\citet{koposov13}. The M giant velocity dispersion presented here
matches within the error-bars the trend of growing trailing tail
velocity dispersion around $\tilde{\Lambda}_\odot = 200^\circ$.

\section{Conclusions}

We have presented a new method to identify M giant stars using
broad-band infrared photometry, more precisely a combination of 2MASS
and WISE data. Our selection procedure picks out stars that are red in
the near-infrared J$-$K colour and blue in the mid-infrared W1$-$W2
colour. Thanks to the presence of the gravity sensitive CO bands in
WISE \change{(leading to a larger colour separation of dwarfs and
  giants)}, the resulting WISE$\cap$2MASS sample of M giants suffers
significantly less contamination \change{(up to a factor of 6)} from
foreground dwarfs compared to the conventional near-infrared
selection.

Taking advantage of the high efficiency of our method, we have been
able to trace Sgr tidal debris at Galactic latitudes as low as
$|b|\sim9^{\circ}$. We have followed our WISE$\cap$2MASS M giant
candidates spectroscopically with AAOmega at the AAT, and show that
the identification success rate is as high as 80\%. Using bona fide M giant tracers,
we provide a new kinematic detection of the Sgr trailing tail debris
at $\tilde \Lambda_{\odot} = 204^{\circ}$. Before crossing the
Galactic disk, the trailing tail appears to be moving faster than
forecasted by the model of \citet{law10}, in agreement with
the prediction of \citet{belokurov14}. 

Our measurement carries important implications for both the disruption
of the Sgr dwarf and the Galactic potential: there is now very little
doubt that the distant tidal debris discovered in the Northern
hemisphere \citep[see, e.g.,][]{DrakeSgr, belokurov14} is indeed the
continuation of the Sgr trailing tail. Such behavior of the trailing
tail implies low differential orbital precession of the Sgr stream,
which in turn could require diminished dark matter content within 100 kpc
\citep[see][]{Gibbons2014}.

\section*{Acknowledgments} 
The research leading to these results has received funding from the
European Research Council under the European Union's Seventh
Framework Programme (FP/2007-2013) / ERC Grant Agreement n. 308024. VB
acknowledges financial support from the Royal
Society. SK acknowledges financial support from the STFC and the
ERC.  EO acknowledges support from NSF grant AST-1313006. DBZ acknowledges the financial support of the Australian
Research Council. This work has made use of the Q3C open source plugin
for PostgreSQL database \citep{koposov06}, Astropy software \citep{astropy2013}, Matplotlib software \citep{Hunter2007}. We also thank the anonymous referee for comments which helped us to improve the paper.

\end{document}